\documentclass[preprint,eqsecnum,aps,showpacs]{revtex4}
\usepackage{dcolumn}
\usepackage{graphicx}
\usepackage{latexsym}
\usepackage{amsmath,amssymb}

\begin{document}

\title{Stable two--brane models with bulk tachyon matter} 

\author{A. Das \footnote{Electronic address: {\em adphy2008@gmail.com}}${}^{}$ and 
Sayan Kar\footnote{Electronic address: {\em sayan@cts.iitkgp.ernet.in}}${}^{}$}
\affiliation{Department of Physics and Centre for Theoretical Studies  \\
Indian Institute of Technology, Kharagpur 721 302, India}            

\author{Soumitra SenGupta \footnote{Electronic address: {\em tpssg@iacs.res.in}}${}^{}$} 
\affiliation{Department of Theoretical Physics  \\
Indian Association for the Cultivation of Science, Kolkata 700 032, India}            

\begin{abstract}
We explore the possibility of constructing stable, warped two--brane models
which solve the hierarchy problem, with
a bulk non--canonical scalar field (tachyon matter) as the source
term in the action. Among our examples are two models--one with a
warp factor (denoted as $e^{-2f(\sigma)}$) which differs from that of
the standard 
Randall--Sundrum
by the addition of a quadratic piece in the $f(\sigma)$ and another, where
the warping is super-exponential. 
We investigate the issue of resolution of hierarchy and
perform a stability analysis by obtaining the
effective inter-brane potentials, in each case. 
Our analysis reveals that
there does exist stable values of the modulus
consistent with hierarchy resolution in both the models. 
Thus, these models,
in which the bulk scalar field generates the geometry and also ensures 
stability, provide viable alternatives to the standard Randall--Sundrum 
two-brane scenario.
\end{abstract}


\pacs{04.50.+h, 04.20.Jb, 11.10.Kk}
\maketitle

\section{Introduction}

It is almost a decade since Randall and Sundrum (RS) \cite{rs2}, following
earlier work \cite{akama,rubakov,visser,add1}, proposed a five dimensional 
model with a warped extra dimension, in order to achieve a resolution of the
hierarchy problem in high energy physics and also to suggest an
alternative to the largely ambiguous procedure of compactification,
necessary in any theory in a spacetime with more than four dimensions.
The RS two--brane model had a positive and
a negative tension brane and we are supposed to live on the one with
negative tension. Such negative tension branes, in general, are not stable. 
Moreover, it was realized that the modulus (representing
the separation between the two 3--branes) in the RS model 
is not {\em stable} and
it is easy to show that the branes would collapse onto
each other due to this instability. In order to rescue this otherwise
elegant model, Goldberger and Wise {\cite{goldwise}} came up with a solution 
using bulk fields, which rendered the model stable. 
The visible brane tension however continued to be negative. 
Different aspects
of the Goldberger--Wise  mechanism have been discussed in numerous articles \cite{dmssgss1,dmssgss2,kg,bgjg,
cjms,jlls, mmsv,bht,alrs}. The stability of Horava--Witten spacetimes
(the work of Horava and Witten {\cite{hw} is the string theoretic
motivation for RS models) was discussed in \cite{lss}.
The Goldberger--Wise (GW)
resolution however assumed a negligible back reaction of the bulk scalar on the metric. However, it was realised
that one could actually solve for the warp factor in the presence
of the cosmological constant and the scalar. The back-reacted warp factor
had a linear and an exponentially decaying piece and by adjusting parameters
one could justify the exclusion of this term and work with the RS one.  
Later, it was also realised that bulk non--canonical bulk scalars (sometimes 
motivated
from string theory) could rescue RS and provide stability. Examples along these
lines were suggested and worked out in \cite{dmssgss1,dmssgss2}. 

In this article, we provide new examples of models with 
a non-canonical bulk scalar and without a cosmological constant. 
More precisely the bulk action is the tachyon matter (or scalar Born-Infeld) 
action\cite{sen1}. 
With this assumption for the matter in the bulk, 
it was shown in {\cite{rksk2}} that one could construct
a viable braneworld model. In Section II, 
we discuss our models in detail. Section III discusses the hierarchy
problem and its resolution in both the models. 
In Section IV, we investigate the stability issue and obtain the
inter-brane potentials for
each model. We also make a comparison between
the two models in this section.
Finally, in Section V, we conclude with some remarks on future
directions.

\section{The models}

The braneworld models we consider in this section are, generically, two brane 
models. However, unlike the RS1 two-brane scenario where there is only
a bulk cosmological constant, we have, in our examples, bulk matter.
In particular, as mentioned in the Introduction, we consider situations 
where the bulk matter is
generated by a non-canonical action, namely that of 
tachyon matter. One of these models have been partially analysed 
in \cite{rksk2}. 
The main features of the
model construction process is, as given below.

We begin with the action for the bulk scalar field given by 
\begin{equation}
S_{T} = \alpha_{T} \int d^{5}\xi \sqrt{-g} V(T)\sqrt{1+g^{MN}\partial_{M} T\partial_{N} T}
\end{equation}
where $\alpha_{T}$ is an arbitrary constant, $g_{MN}$ being the
five dimensional metric. The scalar field is represented by T and 
V(T) corresponds to its potential. The constant $\alpha_{T}$ can take
either positive or negative values. 

The full action for our model is
\begin{eqnarray}  
S & = & S_{G} + S_{T} + S_{B} \\
S_{G} & = &
2 M^3 \int \sqrt{-g} R d^{5} \xi \\
S_{B} & = & - \int  \sum_{j}  \tau_{(j)} \sqrt{-\tilde{g}^{(j)}} d^{4} \xi 
\end{eqnarray}

where $\tilde{g}_{\mu\nu}^{(j)}$ is the induced metric on the brane and
$\tau_{(j)}$ is the vacuum energy density (brane tension) of the $j$th brane. 
There are two branes in a RS-I set-up and a single brane in
RS-II set-up. The $S_B$ terms incorporate the branes and the $S_G$
term takes care of five dimensional gravity. In addition, there is the
matter part $S_T$.  
Variation of the total action w.r.t. 
$g_{MN}$ leads to the Einstein--scalar equations and that w.r.t. T 
gives rise to the scalar field equation.

The five dimensional bulk metric with four dimensional Poincare invariance is 
taken to be of the form: 
\begin{equation}
ds^2 = e^{-2 f(\sigma)}\eta_{\mu\nu} d\xi^{\mu}d\xi^{\nu} + d\sigma^2
\end{equation}
Here $\sigma$ denotes the extra dimensional coordinate and $\xi$ are
the coordinates on a $\sigma=$constant hypersurface.
The warp factor $(f)$ and the scalar field $(T)$ are 
considered to be functions of $\sigma$ only. The extra dimension is 
compactified on an $S^1/Z_2$ orbifold and we assume the two branes are 
located at the fixed points of the orbifold, $\sigma=0,\, \sigma=\pi r$. 
With these
ansatze, the Einstein-scalar equation without a cosmological constant 
reduces to the following system of
coupled, nonlinear ordinary differential equations :
\begin{eqnarray}
{f'}^2 & = & \frac{a}{2} \frac{V(T)}{\sqrt{1 + T'^2}} 
\\
f '' & = & - a V(T) \frac{T'^2}{\sqrt{1 + T'^2}} + \frac {1}{12 M^3} \sum_{j}\tau_{j}
\delta(\sigma - \sigma_{j}) 
\end{eqnarray}
where $a =  \frac{\alpha_{T}}{12 M^3}$ and a prime denotes derivative with
respect to the coordinate $\sigma$. We choose $\alpha_T=1$ henceforth. 
In general, we do not expect to obtain
a general solution of the above equations. 
However, we will discuss two special solutions of these equations below.
A look at the equation will immediately suggest that, in the bulk (i.e.
excluding the brane terms) the relation $\frac{f''}{{f'}^2}=-2 {T'}^2$
holds. Thus, postulating T one can obtain $f$. Further, we can multiply 
the LHS and RHS of the two equations and obtain an expression for
$[V(T)]^2$, which depends on $f$, $T$ and their derivatives. We follow
this approach while obtaining solutions. Note however that a $f(\sigma) \sim
\vert \sigma \vert$ would result in a constant T in the bulk (away from
the branes), since $f''$ would be equal to zero everywhere in the bulk.
This will imply a triviality because the tachyon matter action would
then become equivalent to having a cosmological constant contribution
in the bulk.  

In addition to the bulk solution, one also needs to look at the
equation of motion for the scalar field. This, for bulk tachyon
matter, turns out to be:
\begin{equation}
\frac{V(T)}{(1+T'^2)} T'' + 4 \frac{f'V(T)}{\sqrt{1+T'^2}} T'-\sqrt{1+T'^2} \frac{\partial V}{\partial T} +\sum_i \frac{\partial \tau_i}{\partial T}\delta (\sigma-\sigma_i)=0
\end{equation}
Integrating this equation once would give the boundary (jump) conditions
on $T'$, given as
\begin{equation}
\left[ T'\right ]_i = -(\frac{(1+T'^2)^{\frac{3}{2}}}{V(T)})_i \left \{\frac{\partial \tau_i}{\partial T} \right \}_i
\end{equation}
where $i$ denotes the location of the i-th brane.

\subsection{Model 1:}

It is possible to construct exact solutions of the above equations
and boundary conditions.
For example, one may choose:
\begin{equation}
f(\sigma) = k_1 \vert \sigma \vert - k \vert \sigma \vert \left (\frac{\vert
\sigma \vert}{\vert \sigma_0 \vert} -1\right )
\end{equation}
Here the warp factor at $\sigma=0$ has a value equal to one
while at $\sigma=\sigma_0=\pi r$ its value is the same as that in the RS model.
Notice that we have a parameter $k_1$ which, as we shall see below,
may be assumed to depend on the modulus $r$.

The full solution of the Einstein equations, i.e. $T(\sigma)$ and
$V(T)$ can be found easily. These turn out to be:
\begin{equation}
T(\sigma)= -\frac{1}{2} \sqrt{\frac{\sigma_0}{k}} \ln \left ((k_1+k)\sigma_0
-2 k \vert \sigma \vert \right )
\end{equation}
\begin{equation}
V(T)= \frac{2}{a\sigma_0^2}\left [ e^{-4\sqrt{\frac{k}{\sigma_0}}T}+k\sigma_0
\right ]^{\frac{1}{2}} e^{-2\sqrt{\frac{k}{\sigma_0}} T}
\end{equation}
The brane tensions are:
\begin{eqnarray}
\tau_{(1)} & = & 24 M^3  \left (k_1+k  \right ) \\
\tau_{(2)} & = & - 24 M^3  \left ( k_1-k \right )
\end{eqnarray}
Interestingly, here both the tensions can be positive if $\frac{k_1}{k}\leq 1$
whereas if $\frac{k_1}{k}\geq 1$ one of them is negative. We shall see
later that requirements of stability and hierarchy resolution imply
that $\frac{k_1}{k}>1$. The tachyon 
field here behaves logarithmically with $\sigma$ and the tachyon potential
has a complicated, but primarily exponential relation with $T$ (which, in
a sense, is somewhat reminiscent of the nature of the tachyon potential in the
context of String theory).

We now need to make use of the boundary condition on the scalar field at
each brane. Choosing the brane tensions as:
\begin{equation}
\tau_i = \pm \frac{A}{2}\sqrt{\frac{\sigma_o}{k}}\left [ e^{-x_i}-\sqrt{b}\arctan\frac{e^{-x_i}}{\sqrt{b}} + C_i\right ]
\end{equation}
where $x_i=2\sqrt{\frac{k}{\sigma_0}} T(\sigma=\sigma_i)$, $b=\sigma_0 k$,
$A=\frac{2}{\sigma_0 a}\sqrt{\frac{k}{\sigma_0}}$ and $C_i$ are integration
constants, which, as we will see, will play a crucial role. The expression with a $+$ sign in 2.15 is for
the brane at $\sigma=0$, while the one with a $-$ sign in 2.15 is for
 the $\sigma=\sigma_0$ brane.
 With the above
choices, we obtain the following relations by identifying the tensions
given above with the ones obtained earlier (i.e.
$\tau_{(1)}  =  24 M^3  \left (k_1+k  \right )$ and
$\tau_{(2)}  =  - 24 M^3  \left ( k_1-k \right )$).
\begin{eqnarray}
C_1= (k_1+k)\sigma_0 +{\sqrt{b}} \arctan \frac{(k_1+k)\sigma_0}{\sqrt{b}}  
  \\
C_2= (k_1-k)\sigma_0 +{\sqrt{b}} \arctan \frac{(k_1-k)\sigma_0}{\sqrt{b}}  
\end{eqnarray}
As we shall see later, the possible value of $k_1\sigma_0$ (denoted as
$z$) is
dictated by the requirement of hierarchy resolution. The allowed
value of $k\sigma_0$ is found from the stability requirements.
These values can then be used to finally evaluate $C_1$ and $C_2$
such that the boundary conditions on the scalar field
hold. Note that setting either or both of $C_1$, $C_2$ to zero
leads to inconsistencies.

 
\subsection{Model 2:}
One can easily verify that another solution
exists with the following expressions for $f(\sigma)$ and $T(\sigma)$
\begin{eqnarray}
f(\sigma) = \frac{a_{1}}{k} \left (1-e^{- k \vert \sigma \vert}\right ) \\
T(\sigma) = \sqrt{\frac{2}{k a_{1}}} e^{\frac{k}{2} \vert \sigma \vert}
\end{eqnarray}
with the potential $V(T)$ given as:
\begin{equation}
V(T) = \frac{8}{ak^2} \frac{1}{T^4} \left (1 +
    \frac{k^2}{4} T^2 \right)^{\frac{1}{2}}
\end{equation}
The brane tensions are:
\begin{eqnarray}
\tau_{(1)} & = & 24 M^3 a_{1} \\
\tau_{(2)} & = & - 24 M^3 a_{1} e^{- k \pi r}
\end{eqnarray}
Notice that the brane tensions are not equal and, one of them is
exponentially smaller (though negative in value) than the other.
Also, the warp factor is decaying and has a value equal to 1
on the positive tension brane. Its value on the negative tension
brane is $e^{-2\frac{a_1}{k}(1-e^{-k \pi r})}$. The tachyon field
grows larger as $\sigma$ increases. The tachyon potential,
is nonsingular in the domain in which the tachyon field is defined. 
Moreover to have appropriate warping from Planck to TeV scale, both 
$a_1$ and $k$ must be positive.

Let us now analyse the boundary conditions on the scalar field T.
Following the method outlined for Model 1, we first write
down our choices for the brane tensions in terms of T.
These are given as:
\begin{equation}
\tau_i = \pm \frac{2k}{a} \left [ \frac{1}{2 x_i^2} +\ln x_i -\frac{1}{2}\ln (1+x_i^2) + D_i\right ]
\end{equation}
where $x=\frac{kT}{2}$ and $D_i$ are integration constants (the parallel
of the $C_i$ in Model 1). The expression with a $+$ sign in 2.15 is for
   the brane at $\sigma=0$, while the one with a $-$ sign in 2.15 is for
   the $\sigma=\sigma_0$ brane. 
Evaluating the tensions at $\sigma_i$ and
equating them to the ones obtained earlier in this subsection, we arrive 
at the following relations.
\begin{eqnarray}
D_1 = \frac{1}{2}\ln{\left (2 y+1\right )} \\
D_2 = \frac{1}{2}\ln \left ( 2 y e^{-x}+ 1\right ) 
\end{eqnarray}
where $y=\frac{a_1}{k}$ and $x=\pi k r$.

The first of the above relations defines $D_1$ for given $y$. The second defines $D_2$
for given $x$ and $y$. As shown later, we get $y$ from hierarchy resolution
and $x$ from stability requirements.

\section{The resolution of hierarchy in these models}

We now turn towards analysing the question of hierarchy resolution in these
models.

\subsection{Model 1}
In this model, the mass scales on the two branes are related as 
\begin{equation}
\frac{m}{m_0}=e^{- \pi k_1 r}
\end{equation}
Using the definitions, $z=\pi k_1 r$, the resolution of hierarchy requires, 
$e^{-z} \sim e^{-40}$ or $z \sim 40$ (or $k_1 r\sim 13$, similar to the
RS model). 

Furthermore, the relation between $M_{Pl}^2$ and $M^3$ turns out to be:
\begin{equation}
M_{Pl}^2 = \frac{M^3 \sqrt{\pi}}{2k} \frac{z}{p} e^{-\frac{1}{4}\frac{z}{p}
(p+2)^2}\left[ Erf\left[ \frac{1}{2}\sqrt{\frac{z}{p}} (p+2)\right ] - Erf \left[ \frac{1}{2}\sqrt{\frac{z}{p}} (p-2) \right ] \right ]
\end{equation}
where $p=\frac{k_1}{k}$ and $Erf$ denotes the error function.

If $z \sim 40$ and $p \sim 40$, we find that $M_{Pl}^2 = 0.024 M^3/k$. 
Therefore
assuming, $M \sim M_{Pl}$ leads to $k\sim 0.24 M_{Pl}$, unlike $M_{Pl}\sim M \sim k$ in the RS model (based on the relation $M_{Pl}^2 \sim \frac{M^3}{k} \left (1-e^{-2k\pi r_c}\right )$ in RS and for moderate $kr_c \sim 12$). Note however, that
the $k$ in our model is not the same as the RS `k'--which is attached to the 
linear part in the RS warp factor. In our model, the parallel of the RS `k' is $k_1$.  
Since $k_1= k p \sim 40 k$ we get $k_1 \sim 0.96 M_{Pl}$. Thus, one can
set $k_1$, $M$ and $M_{Pl}$ of the order of $M_{Pl}$ while $k$ needs to
be set at an order of ${10}^{-2} M_{Pl}$. Therefore, in this model,
an extra small hierarchy, somewhat similar to the ADD scenario exists 
\cite{add1} and cannot be avoided.

\subsection {Model 2}
We note that the relation between mass scales on the branes at $\sigma=0$ and 
$\sigma=\pi r$ is
\begin{equation}
m= m_0 e^{-f(\pi r)}=e^{-y(x) \left (1-e^{-x}\right )}
\end{equation}
To solve the hierarchy problem one needs $\frac{m}{m_0} \sim e^{-40}$. Thus,
$\frac{a_1}{k}$ must be around 40. 

In addition, the relation between $M_{Pl}^2$ and $M^3$ turns out to be:
\begin{equation}
M_{Pl}^2= \frac{2 M^3}{k} e^{-2y}\left [ Ei[2 y]-Ei[2 y e^{-x}] \right ]
\end{equation}
where $Ei$ denotes the exponential integral function.

With $y=40$ and $x$ moderate (about 3 or larger, as we will see from the
stability analysis), we obtain, for $M_{Pl} \sim  M$, $k \sim 0.025 M_{Pl}$. 
As in the Model 1,
we have the parameter $a_1$ as the parallel of the RS `k' and this takes
on the value $a_1 \sim 40 k \sim M_{Pl}$. Similar to Model 1, an extra
small hierarchy seems to exist here too (through the value of the k in our model)
and the resolution of hierarchy seems to
have an ADD-like flavour \cite{add1}. 

\section{Stability analysis: inter-brane potentials}

We shall now investigate the stability of our two-brane models.

Let us first look at the original RS-I model (with only a bulk
negative cosmological constant) and construct the interbrane potential.
This turns out to be (omitting the $\int d^4 \xi$) 
\begin{equation}
V(x) = -18 M^3 k \left ( 1- e^{-4 x}\right )
\end{equation}
where $x=\pi k r$ and the warp factor is just $e^{-k\vert \sigma \vert}$.
Note that there is only one free parameter here (i.e. $x$) and the
potential does not have any minimum.

This potential will be modified, if we also take into account
the five-dimensional gravitational part of the  action and integrate
it over the extra dimension. In summary, the gravity part is given as
$\int \sqrt{g} {}^5 R d^5 \xi$  which has a piece which would give
4D gravity upon integrating over extra dimensions (provided we have
a curved brane) and another piece
(proportional to $\int e^{-4f} \left (8 f''-20 {f'}^2\right ) d\sigma d^4 \xi$).
It is a debatable issue as to whether we should 
include this piece in the potential or not. We, however remain non-commital
on this and discuss both scenarios (i.e. with and without the gravitational
piece).
Including this so--called gravity part into the inter-brane potential
for RS-I,
we find that
\begin{equation}
V(x) = 4 M^3 k \left (1- e^{-4 x}\right )
\end{equation}
The potential is now positive definite but, as before, has no minimum.

The above analysis shows that, at least for RS-I, by including the gravity part one cannot make the branes stable. Goldberger and Wise \cite{goldwise} therefore
included an extra bulk scalar field which was crucial in order
to achieve stability. 
  
In our models, we have bulk tachyon matter which generates the spacetime
geometry. In both the models discussed above we have two parameters(
z, p in Model 1 and x, y in Model 2).  
Following GW, we consider the matter action (tachyon--matter) + 
the two brane actions and insert the
solutions for $T(\sigma)$ and $f(\sigma)$ in them. Thereafter, we integrate
over the extra dimension to obtain the effective inter-brane potential as a
function of $r$.

As mentioned before, in either model, we have two independent parameters and, 
further, a pair of constants
which depend on these parameters. We adopt the following strategy to
find the allowed values of the parameters and the constants.
Let us assume that in either model the parameters are related 
functionally. We do not provide any compelling reason for doing so--however
there is no reason either why such a relationship may be disallowed.
In other words, in Model 1, we have $p(z)$ and in Model 2,
$y(x)$. We can now choose simple forms of these functions (preferably
trigonometric) in order to have a
minimum in either potential. Let us now focus on the two models
separately.    

\subsection{Model 1:}

Firstly, let us follow the procedure outlined above to obtain the potentials. 
Define $p=k_1/k$ as before, so that
\begin{equation}   
f(\sigma)=k\left[p \vert \sigma \vert -\vert \sigma \vert \left(\frac{\vert \sigma\vert}{\vert \sigma_0 \vert}-1 \right) \right]
\end{equation}
Now let $z=\pi k_1 r$ and $p=p(z)$. Then, the expressions for the potentials, 
including and excluding the Ricci scalar contribution in the action are,
respectively (omitting the integration over $d^4 x$),

\noindent {\sf Without gravity term contribution:}
\begin{eqnarray}   
V(z)&=&-24M^3k(p+1) + 24M^3 k (p- 1)e^{-4z}+\int_0^1 dx\,\,24M^3 k e^{-4z(x + x/p - x^2/p)}\nonumber \\ 
&&+\int_0^1 dx\,\, 24M^3 kz p e^{-4z(x + x/p - x^2/p)}(1 + 1/p - 2x/p)^2 
\end{eqnarray}

\noindent {\sf With gravity term contribution:}
\begin{eqnarray}
V_{mod}(z)&=&8M^3k(p+1)-8M^3k(p-1)e^{-4z}-\int_0^1dx \,\, 8M^3ke^{-4z(x+x/p-x^2/p)} \nonumber \\ 
&& - \int_0^1dx \,\, 16M^3zk_1e^{-4z(x+x/p-x^2/p)}(1+\frac{1}{p}-\frac{2}{p}x)^2
\end{eqnarray}

where, in the above integrals $x=\frac{\sigma}{\sigma_0}$. Note, both
the integrals above can be done--we do not write them out here
because the expressions are complicated and it is more useful to plot
the potentials directly. It should also be realised that the dominant
contributions to the potentials essentially come from the first two terms
in each of them.

The strategy to find possible stable values of z will be as follows.
Note that the hierarchy resolution requires $z=40$. Let us choose
$p(z)=p_0 {\cos^2 z}$. This restricts $p$ to have values between $0$ and $p_0$.
Further it ensures the presence of minima in $V(z)$ since the
{\em linear in p} terms will contribute to their existence. Using this
form of $p(z)$ (with $p_0=40$, say) 
in either of the above expressions
for the potentials we find that there is a minimum around $z=40$ (the value
required for hierarchy resolution). 
We show the potentials in Fig. 1 and Fig. 2.
Since $z$ and $p$ values are now known and we can easily find out 
$C_1$ and $C_2$
by substituting these values in the expressions for the $C_i$ given earlier.

\begin{figure}[tbh!]
\includegraphics[width=7cm,height=7cm]{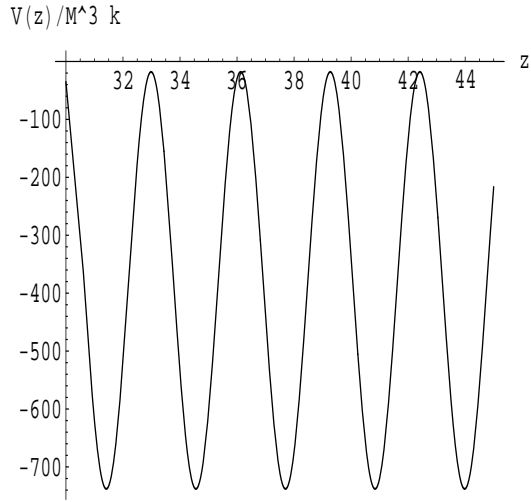}
\caption{Potential $V(z)/M^3k$ vs. $z$ (without gravity term contribution) for Model 1. Note a minimum at $z\sim 41$.}
\label{sec_warp_1}
\end{figure}
\begin{figure}[tbh!]
\includegraphics[width=7cm,height=7cm]{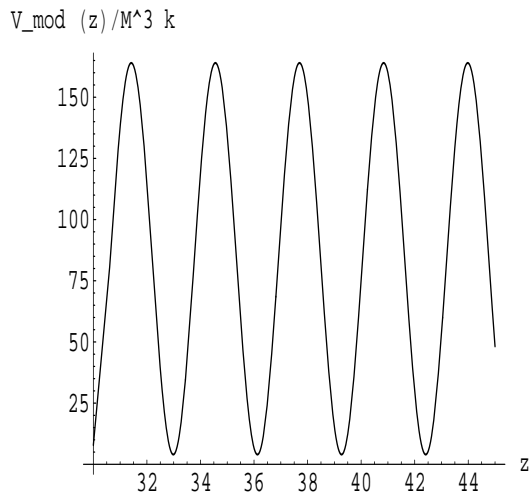}
\caption{Potential $V_{mod}(z)/M^3k$ vs. $z$ (with gravity term contribution) for Model 1. Note a minimum around $z\sim 39.5$.}
\label{sec_warp_2}
\end{figure}

\subsection{Model 2:}

For the analysis of the stability of branes in Model 2,
recall, the dimensionless variables $y=\frac{a_1}{k}$ and
$x=\pi k r$. Omitting, as before, the explicit writing of the integration 
over $d^4 \xi$ we obtain the following effective potentials:

\noindent {\sf Without gravity term contribution:}

\begin{equation}
V(x)=  M^3 k \left [ \frac{3}{2} e^{-4 y} e^{4 y e^{-x}} \left (12 y e^{-x}-1\right ) - 18 y + \frac{3}{2} \right ]\label{v_old}
\end{equation}

\noindent {\sf With gravity term contribution:}

\begin{equation}
V_{mod}(x) = 4M^3ky\left[1-e^{-4y}e^{-x}e^{4ye^{-x}} \right]
\end{equation}

The dominant contribution in each potential comes from the {\em linear in y}
term.
In either expression, we now choose $y(x)=y_0 \cos^2 x$ (with $y_0\sim 40$). 
Once again, this
choice restricts $y$ to values between $0$ and $40$. With this
choice, we now plot the potentials and see whether there are
minima at finite values of x.

The figures (Fig. 3 and Fig. 4) below demonstrate that we do indeed have minima in the potentials
such that $y$ is close to $40$ at some finite positive $x$.

\begin{figure}[htb!]
\includegraphics[width= 10cm,height=6cm]{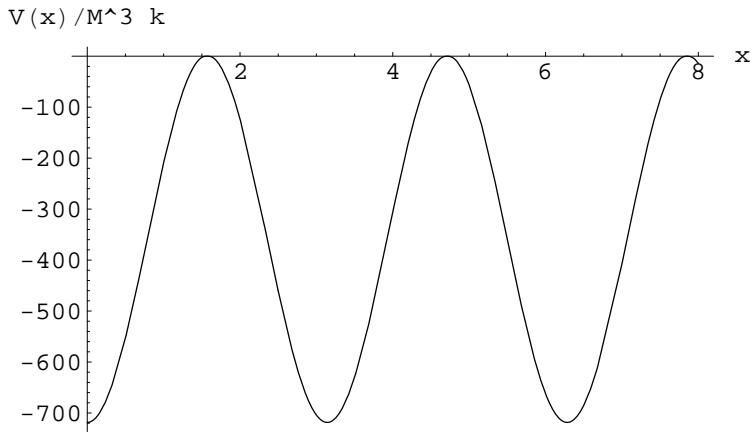}
\caption{Plot for $V(x)/M^3k$ vs. x (without gravity term contribution) for Model 2. Note a minimum around $x\sim 3$.}
\end{figure}

\begin{figure}[tbh!]
\includegraphics[width=5cm,height=5cm]{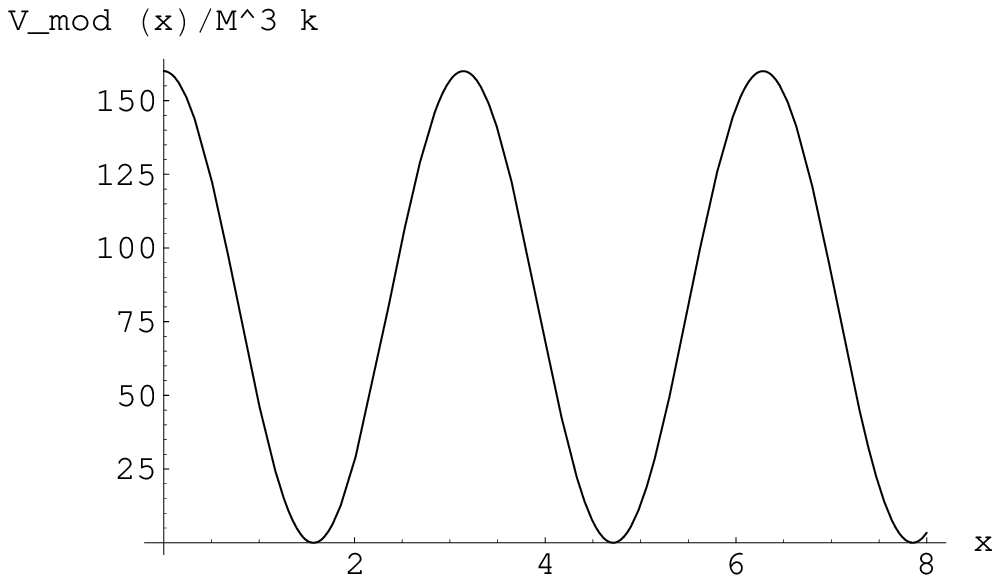}
\caption{Plot of $V_{mod}(x)/M^3k$ vs. $x$ (with gravity term contribution) for Model 2. Note a minimum around $x\sim 4.7$.}
\end{figure}

Thus, as in Model 1, knowing $x$ and $y$ we can easily find out $D_1$ and $D_2$
from the abovementioned expressions.

\subsection{Comparison between the two models}

It is reasonable to ask--which among the above two is a better model. 
A very clear answer to this question does not seem to exist, at this stage. 
We list below a few points
which might shed light on this aspect.

$\bullet$ The second model is very different from RS in terms of the
functional form of the warp factor. The first one can, in a way, be thought of 
as a perturbation.

$\bullet$ At the level of solutions (without invoking any constraints
that arise from hierarchy resolution or stability), Model 1
seems to allow the existence of two positive tension branes, while
Model 2 does not. This feature of Model 1 differs from what we
obtain in standard RS where there is no scope of having two
positive tension branes even at the level of solutions. 
 
$\bullet$ In the first model the scaling of masses occurs in a way
different (super--exponential) from RS whereas in the second model the 
scaling is the same (exponential) as for RS.

$\bullet$ The nature of the tachyon potential in the first model is
far removed from whatever string theorists write down from varied
considerations (which, in some sense may not be applicable here because
there arguments are based on cosmological considerations--rolling tachyon
etc.). In the second model, by virtue of being exponential (not exactly,
but in some way related) the tachyon potential is perhaps a bit closer to
the stringy ones.

$\bullet$ The stability issue is solved in both the models by incorporating
a similar logic (i.e. a functional realtionship between the two
dimensionless parameters). Neither model can be called as preferred, from this
perspective.

$\bullet$ For hierarchy resolution, we have seen that Model 2 is
different in nature than Model 1, which is closer to RS. 
This is because in Model 1, 
the value of $p$ does not affect
hierarchy resolution as long as the choice of $p(z)$ results in a
minimum in $V(z)$ at the right value of $z$ (required for the solution
of the hierarchy problem). On the other hand, in Model 2, 
a proper value of $y$ is
required for hierarchy resolution and the function $y(x)$ should attain
that value at a minimum in the potential. 

$\bullet$ Further, in the resolution of hierarchy in either model we
seem to be introducing an element of hierarchy through the values of
the extra parameters (the $k$ in Models 1 and 2). This aspect seems to be closer
to the hierarchy resolution in the ADD model \cite{add1}, though it is not, 
in any way dependent on the number of extra dimensions, as in ADD 
where it happens through the
factor associated with the volume of the extra dimensions.

In order to facilitate a comparison we tabulate the essential results
of Models 1 and 2 in Table I. 

\vspace{0.2in}
\begin{center}
\begin{tabular}{|c|c|c|}\hline \hline
{\bf Feature} & {\bf Model 1} & {\bf Model 2} \\ \hline\hline 
{Warp factor} & $f(\sigma) = k_1 \vert \sigma \vert - k \vert \sigma \vert \left (\frac{\vert
\sigma \vert}{\vert \sigma_0 \vert} -1\right )$
  & $f(\sigma) = \frac{a_{1}}{k} \left (1-e^{- k \vert \sigma \vert}\right )$  \\ \hline
{Tachyon field} & $T(\sigma)= -\frac{1}{2} \sqrt{\frac{\sigma_0}{k}} \ln \left ((k_1+k)\sigma_0
-2 k \vert \sigma \vert \right )$
 & $T (\sigma) = \sqrt{\frac{2}{k a_{1}}} e^{\frac{k}{2} \vert \sigma \vert} $ \\ \hline
{Tachyon Potential} & $V(T)= \frac{2}{a\sigma_0^2}\left [ e^{-4\sqrt{\frac{k}{\sigma_0}}T}+k\sigma_0
\right ]^{\frac{1}{2}} e^{-2\sqrt{\frac{k}{\sigma_0}} T}$
 &  $V(T) = \frac{8}{ak^2} \frac{1}{T^4} \left (1 +
    \frac{k^2}{4} T^2 \right)^{\frac{1}{2}}$
 \\ \hline
{Parameters} & k, $k_1$, r & $a_1$, k, r \\ \hline
{Brane tensions} & $\tau_{(1)} = 24 M^3  \left (k_1+k  \right )$,
 & $\tau_{(1)}  =  24 M^3 a_{1}$ \\
& $\tau_{(2)} = - 24 M^3  \left ( k_1-k \right )$ &
$\tau_{(2)}  = - 24 M^3 a_{1} e^{- k \pi r}$ \\ \hline
{Reduced parameters} & $z=\pi k_1 r$, $p=\frac{k_1}{k}$ & $y=\frac{a_1}{k}$, $x=\pi k r$  \\ \hline
{B.C. and constants} & $C_{1,2}$ (Eq. 2.16, 2.17)  &  $D_{1,2}$ (Eq. 2.24, 2.25)\\ \hline
{Relations} & $p=p_0 \cos^2 z$ & $y=y_0 \cos^2 x$ \\ \hline
{Hierarchy resolution} & $m= m_0 e^{-z}$ & $m= m_0 e^{-y(1-e^{-x})}$ \\
& $m\sim$ TeV, $m_0 \sim M_{Pl}$, $ z\sim 40$ & $m \sim$ TeV, $m_0\sim M_{Pl}$,
$y\sim 40$\\  
 & & $x \sim 3$ to $5$ \\ \hline
{$M_{Pl}$, M  relation} & Eq. 3.2 gives for & Eq. 3.4 gives for \\ 
& $M\sim M_{Pl}\sim k_1$, $k\sim 0.024 M_{Pl}$ & $M\sim M_{Pl}\sim a_1$, $k\sim 0.025 M_{Pl}$ \\ \hline
{Inter-brane potentials} & Eq. 4.4, Eq. 4.5  & Eq. 4.6, Eq. 4.7 \\ \hline
{Stability} & Pot. minimum around $z\sim 40$ & Pot. minimum around $x\sim 3$ to $5$ \\ 
& $p=p_0 \sim 40$, $\cos^2 z\sim 1$ & $y=y_0\sim 40$, $\cos^2 x \sim 1$ \\ \hline 
{Deficiency} & Additional small hierarchy in k  & Additional small hierarchy in k \\ \hline
\end{tabular}
\end{center}
\vspace{0.2in}

\centerline{Table 1: Summary and comparison of the two models.}

\section{Remarks and conclusion}

In this article, we have tried to build two-brane models with bulk
matter, which solve the stability and hierarchy problems.

How different are our models from the standard Randall-Sundrum ?
From Table 1, we can write down the warp factors for the two
models using the values of the parameters quoted there. These
turn out to be:

\newpage

\noindent {\bf Model 1:}

\begin{equation}
f(\sigma)= M_{Pl} \left [ \vert \sigma \vert - 0.024 \vert\sigma \vert 
\left (\frac{\vert \sigma \vert}{40} M_{Pl}-1\right ) \right ]
\end{equation}

\noindent {\bf Model 2:}

\begin{equation}
f(\sigma) = 40 \left (1-e^{-0.025 M_{Pl}\vert \sigma \vert} \right )
\end{equation}
 
If we plot the above two warp factors along with the linear-in-$\sigma$ RS one 
we will see that the deviations are very small and the overall decaying nature
is maintained. So, where does this small difference make a crucial contribution?
To see this let, us go back to the expressions for the inter-brane potentials.
In either model, the inter-brane potentials are dominated by the
linear terms ($z$ in Model 1 and $y$ in Model 2), both of which are
essentially the ratio of each of two parameters ($k_1$, $k$ in Model 1 and
$a_1$, $k$ in Model 2). The existence of these linear terms therefore
depends on the existence of more than one parameter (apart from $r$) 
in either model, which, in turn, is
made possible through the presence of bulk matter. Postulating a relationship
between the parameters (i.e. $y(x)$ or $p(z)$) thus becomes a possibility
(unlike the RS case) and hence shows a way of achieving stability.
  
The tachyon matter scalar field is responsible
for providing the source for the bulk metric and keeping the
branes stable. The crucial point in our analysis has been the
fact that among the two free parameters in each of our models (x,y and z,p),
we have invoked relationships (i.e. $y(x)$, $p(z)$).
This has facilitated the resolution of the stability and hierarchy problems
simultaneously. 
We agree that we do not have a good reason
behind choosing $y(x)$ or $p(z)$ though, as shown, trigonometric functions
seems to be the preferred choice in order to have minima in the potentials
and hence, stability.

The issue of whether to include the contribution of the gravitational
part of the action in defining the effective inter-brane potential
remains an open issue. We have preferred to be non--commital on this
in our paper and that is why we include results for both scenarios.
However, we may note that we do not see any major difference in our
results after including the gravitational part of the action. 
The only difference seems to be the fact the in the presence of
the gravity part the potential in all cases is positive, while, otherwise,
it is negative. If we
had seen a more significant difference (i.e. say, the 
 existence of minima with the gravity part
and no minima without or vice versa), we might have had a case for choosing one 
over the other. At this stage, we do not make any definite statement
on this aspect. 

We conclude this article by stating that our analysis of stability
and our proposal on a functional relationship between the parameters
in our models, surely needs more justification at a more basic level.
In addition, it
is crucial to address field localisation, massive modes, KK graviton
effects, corrections to Newtonian gravity and other phenomenological issues
using the new warp factors introduced in this article.  
We hope to address some of these issues in our future investigations.

\section*{Acknowledgments}

AD thanks CSIR, New Delhi, India  for financial support through a research fellowship.



\begin{thebibliography}{99}

\bibitem{rs2} 
L. Randall and R. Sundrum, Phys. Rev. Lett. {\bf 83} 3370 (1999);
{\em ibid.} Phys. Rev. Lett. {\bf 83}, 4690 (1999)

\bibitem{akama} K. Akama, in {\em Proceedings of the Symposium on Gauge 
Theory and Gravitation}, Nara, Japan (Springer-Verlag, 1982)
hep-th/0001113

\bibitem{rubakov} 
V. A. Rubakov and M. E. Shaposhnikov, Phys. Lett. {\bf B 125}, 139 (1983)

\bibitem{visser} M. Visser, Phys. Letts. {\bf B 159},22 (1985), hep-th/9910093


\bibitem{add1} N. Arkani-Hamed, S. Dimopoulos  and  G. Dvali, 
Phys. Lett. {\bf B429} 263-272 (1998); I. Antoniadis, N. Arkani-Hamed, 
S. Dimopoulos and G. Dvali, Phys. Lett. {\bf B436} 257-263 (1998);
N. Arkani-Hamed, S. Dimopoulos and G. Dvali, Phys. Rev. {\bf D59} 086004-21
(1999)

\bibitem{goldwise} W. D. Goldberger and M. B. Wise, Phys. Rev. {\bf D60},107505
(1999) {\em ibid.} Phys. Rev. Letts. {\bf 83},4922 (1999)

\bibitem{dmssgss1} D. Maity, S. SenGupta, S. Sur, Phys. Lett. {\bf B}643, 348

\bibitem{dmssgss2} D. Maity, S. SenGupta, S. Sur, arXiv:hep-th/0609171; 
A. Dey, D. Maity, S. SenGupta, Phys. Rev. {\bf D75}, 107901 (2007); S. Das, A. Dey, 
S. SenGupta, arXiv:0704.3119 

\bibitem{kg} K. Ghoruku and A. Nakamura, Phys. Rev. {\bf D64}, 084028 (2001);
K. Ghoruku, arXiv:hep-th/0402102

\bibitem{bgjg} B. Grzadkowski and J. F. Gunion, Phys. Rev. {\bf D 68}, 
055002 (2003)

\bibitem{cjms} D. Choudhury, D.P. Jatkar, U. Mahanta, S Sur, JHEP (09), 021 (2000)

\bibitem{jlls} J. Lesgourgues, L. Sorbo, Phys.Rev. {\bf D69} 084010 (2004)

\bibitem{mmsv} A. S. Mikhailov, Y. S. Mikhailov, M. N. Smolyakov, I. P. Volobuev,
 Class.Quant.Grav. {\bf 24} 231 (2007)

\bibitem{bht} F. Bruemmer, A. Hebecker, E. Trincherini, Nucl.Phys. {\bf B738}
283 (2006)

\bibitem{alrs} A. Lewandowski and R. Sundrum, Phys. Rev. {\bf D 65}, 044003 (2002)

\bibitem{hw} P. Horava and E. Witten, Nucl. Phys. {\bf B 460}, 506 (1996);
Nucl. Phys. {\bf B475}, 94 (1996)

\bibitem{lss} J.-L.Lehners, P. Smythe and K. S. Stelle, Class.Quant.Grav. 
{\bf 22} 2589 (2005)

\bibitem{sen1} A. Sen, JHEP 0204,  048 (2002); A. Sen, JHEP 0207, 065 (2002); 
A. Sen, Mod.Phys.Lett. {\bf A}17, 1797 (2002)
 
\bibitem{rksk2} R. Koley, S. Kar, Phys.Lett. {\bf B}623 244 (2005); Erratum-ibid. {\bf B}631 199 (2005)

\end{thebibliography}
\end{document}